\documentclass[prb,twocolumn,superscriptaddress,nopacs]{revtex4}
\usepackage{graphicx}

\usepackage{amsmath}
\usepackage{amssymb}
\usepackage{accents}
\usepackage{color}
\usepackage{verbatim}
\usepackage{hyperref}
\usepackage{empheq} 
\definecolor{very-light-gray}{gray}{0.9}

\usepackage{float} 

\renewcommand{\r}{\mathbf{r}}

\begin{document}

\title{Charge Smoothening and Band Flattening due to Hartree corrections in Twisted Bilayer Graphene}
\author{Louk Rademaker}
\affiliation{Department of Theoretical Physics, University of Geneva, 1211 Geneva, Switzerland}
\author{Dmitry A. Abanin}
\affiliation{Department of Theoretical Physics, University of Geneva, 1211 Geneva, Switzerland}
\author{Paula Mellado}
\affiliation{School of Engineering and Sciences, Universidad Adolfo Ib\'{a}\~{n}ez, Santiago 7941169, Chile}

\date{\today}

\begin{abstract} 
Doping twisted bilayer graphene away from charge neutrality leads to an enormous buildup of charge inhomogeneities within each Moir\'{e} unit cell. Here we show, using unbiased real-space self-consistent Hartree calculations on a relaxed lattice, that Coulomb interactions smoothen this charge imbalance by changing the occupation of earlier identified `ring' orbitals in the AB/BA region and `center' orbitals at the AA region. For hole doping, this implies an increase of the energy of the states at the ${\bf \Gamma}$ point, leading to a further flattening of the flat bands and a pinning of the Van Hove singularity at the Fermi level. The charge smoothening will affect the subtle competition between different possible correlated phases.
\end{abstract}

\maketitle


Studies of twisted bilayer graphene have revealed a number of striking electronic properties,\cite{TramblydeLaissardiere:2010hla,Luican:2011hw,Brihuega:hh,Wong:2015dg,SuarezMorell:2010bz,Bistritzer:2011ho,LopesdosSantos:2012vk,Slotman:2015kd,vanWijk:2015bc,Fang:2016iq,Nam:2017jh} and recently culminated in the discovery of a correlated insulator and superconducting phase.\cite{Cao:2018kn,Cao:2018ff} Several new experiments performed since then\cite{Cao:2019ui,Yankowitz:2018tx,Choi:2019aa,Xie:2019vz,Jiang:2019to,Lu:2019ww,Sharpe:vc} 
have revealed new correlated and topological states, and even more theoretical works have appeared\cite{Xu:2018,Po:2018vk,
Kang:2018wd,Kang:2018tt,Zou:2018ud,Po:2018,Xie:2018wn,Rademaker:2018fy,Pizarro:2018wx,Wu:2018vh,Kennes:2018wi,Yuan:2018un,Efimkin:2018us,Liu:2018ur,Choi:2018vw,Fu:2018vx,PhysRevX.8.031087,Venderbos:2018vs,Tarnopolsky:2018vv,Song:2018ul,Sherkunov:2018wf,Thomson:2018ve,Ochi:2018ug,Zhang:2018wj,Wu:2018w7,Seo:2018ha,Laksono:2018dma,Guinea:2018kd,Cea:2019tb,Guinea:2019ul,Haule:2019wi,Bultinck:2019wp,Leaw:2019vz,Yan:2019vp,Walet:2019um,Chou:2019uf,Gonzalez:2019up,Li:2019wj,Wu:2019nj,Gu:2019vq,Yudhistira:2019fd}. However, there is no consensus yet on the correct low-energy effective band-structure, let alone the relevant many-body interactions.

It is well known that at fillings away from charge neutrality the electronic charge piles up in the AA regions of the Moir\'{e} unit cell.\cite{TramblydeLaissardiere:2010hla,Luican:2011hw,Brihuega:hh,Wong:2015dg,Rademaker:2018fy,Guinea:2018kd,Guinea:2019ul}
The Coulomb interaction will counterbalance the formation of such large charge inhomogeneities, and is expected to significantly modify the non-interacting band dispersion. Here we study the effect of the electronic interactions on AA localization via an unbiased, self-consistent Hartree calculation.

Starting from a microscopic tight-binding model on a relaxed twisted bilayer at the magic angle $\theta = 1.08^\circ$ at fillings at and below charge neutrality, we show that the Hartree corrections indeed cause the redistribution of charges within the unit cell. The resulting charge distribution is much smoother than that found with noninteracting theories. The smoothening itself is made possible by increasing the energy of the states at the ${\bf \Gamma}$ point in the Brillouin zone, which causes a further substantial flattening of the (already nearly) flat bands. Our results are qualitatively similar to earlier results based on a continuum model.\cite{Cea:2019tb,Guinea:2018kd} Given the tight competition between different possible correlated phases,\cite{Xu:2018,Ochi:2018ug,Kang:2018tt} the charge transfer reported here will almost certainly affect this subtle competition.

In the remainder of this paper, we first introduce our model of twisted bilayer graphene and the details of our calculation. We then present and compare the charge inhomogeneities before and after the Hartree corrections. Finally, we discuss the ramifications of our result for future theoretical modelling.

\begin{figure}
	\includegraphics[width=0.49\columnwidth]{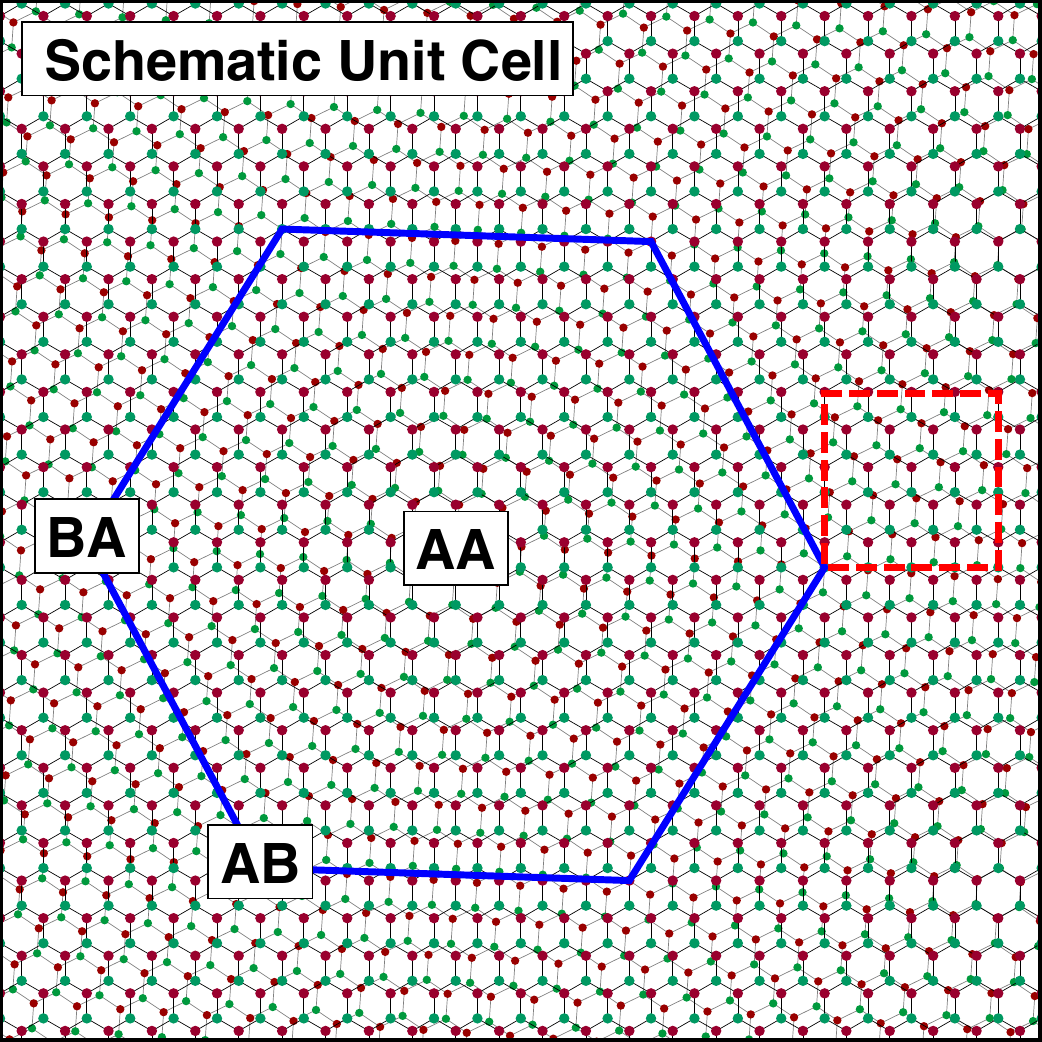}
	\includegraphics[width=0.49\columnwidth]{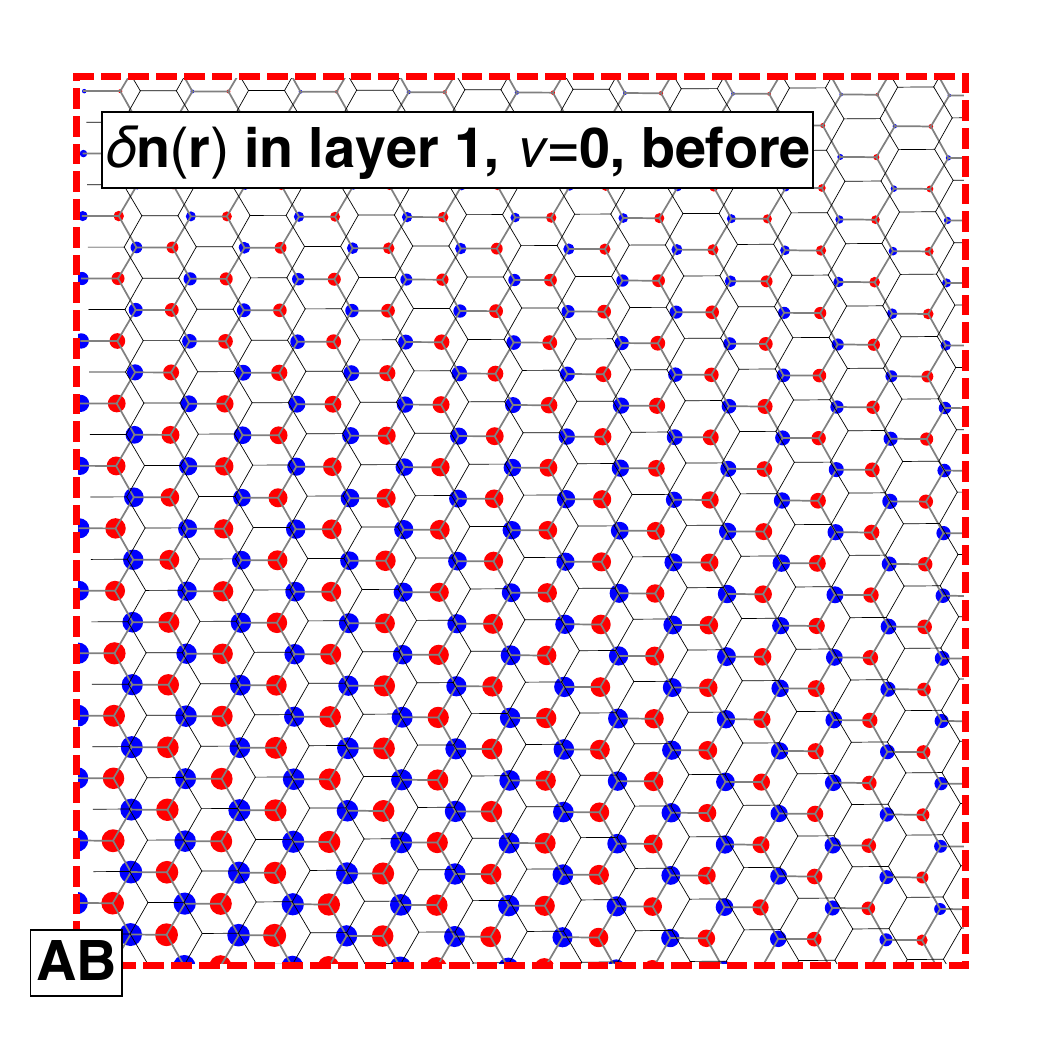}
	\caption{\label{Fig:ABOscillations}
	{\bf Left:} A schematic representation of the Moir\'{e} unit cell in twisted bilayer graphene, with indicated AA, AB and BA regions (a larger twist angle shown). The unit cell itself is shown with a blue hexagon. The red dashed square indicates the region enlarged in the right figure.
	{\bf Right:} The relative electron charge density $\delta n (\r)$ in the AB region of the unit cell at charge neutrality $\nu=0$ in layer 1 before Hartree corrections. Atoms that lie on top of atoms in the other layer have a charge deficiency (blue circles) and the remaining atoms have a charge excess (red circles). The size of the circles indicate the magnitude of $\delta n$, with a maximum of $|\delta n| \sim 0.0025$ at the AB point, here shown in the left bottom corner.
	}
\end{figure}

\emph{The model} -- 
Our starting point is a tight-binding model of twisted bilayer graphene, \cite{Rademaker:2018fy} which builds on earlier continuum models.\cite{Bistritzer:2011ho} 
We construct twisted bilayer graphene by starting with an AB stacked bilayer and rotating the top layer around an AB site. For a commensurate twist angle, we choose two integers $m_1, m_2$ such that the large Moir\'{e} unit cell has unit vectors ${\bf G}_1 = m_1 {\bf a}_1 + m_2 {\bf a}_2$ and ${\bf G}_2 = -m_2 {\bf a}_1 + (m_1+m_2) {\bf a}_2$, 
where the graphene lattice unit vectors are ${\bf a}_{i} = \tfrac{a}{2} \left( (-1)^i \hat{\bf x} + \sqrt{3} \hat{\bf y} \right) $ with $a = 0.246$ nm\cite{Reich:2002kd,CastroNeto:2009cl}.
The twist angle is now given by $\cos \theta = \frac{m_1^2 + 4 m_1m_2 + m_2^2}{2 (m_1^2 + m_1m_2 + m_2^2)}$, with $4 (m_1^2 + m_1m_2 + m_2^2)$ being the number of atoms per unit cell. In the remainder of this paper we focus on $m_1=30$, $m_2=31$, which corresponds to the twist angle $\theta = 1.08 ^\circ$ and $N = 11,164$ atoms in the Moir\'{e} unit cell. A schematic picture of the unit cell is shown in Fig.~\ref{Fig:ABOscillations}, left.

At such small twist angles the atomic positions are substantially changed when one allows for lattice relaxation. To include this effect, we use the relaxed atomic positions from Ref.~[\onlinecite{Gargiulo:2018bj}].\footnote{We thank Prof. Yazyev for kindly providing us with a list of atomic positions in the relaxed lattice.} Using these atomic positions, we define a tight-binding hopping model with hopping between two atoms given by
\begin{equation}
{\scriptstyle
	-t(\vec{d}) = V^0_{\pi} \left[ 1 - \left(\frac{\vec{d}\cdot \vec{e}_z}{d} \right)^2 \right] e^{- \frac{d - a_0}{r_0} }
		+ V^0_{\sigma} \left(\frac{\vec{d}\cdot \vec{e}_z}{d} \right)^2  e^{- \frac{d - d_0}{r_0} }
		}
\end{equation}
Here $a_0 = 0.142$ nm is the intralayer nearest neighbor atomic distance; $V^0_\pi = 2.7$ eV is the nearest neighbor intralayer hopping strength; $d_0 = 0.335$ nm is the unrelaxed interlayer distance; $d = |\vec{d}|$ is the length of vector that connects the two carbon atoms; $r_0 = 0.045$ nm is the $p$-orbital decay length; and $V^0_\sigma = 0.48$ eV is the interlayer hopping when two carbon atoms are exactly above each other. \cite{Nam:2017jh,TramblydeLaissardiere:2010hla}.

The electrons are subject to Coulomb interactions, which depend on the deviation of the electron density $n(\r)$ from the average density $\bar{n}$. With $\delta n(\r) \equiv n(\r) - \bar{n}$ the interaction Hamiltonian reads
\begin{equation}
	H_{V} = \frac{1}{2} \sum_{ij} 
		\delta n(\r_i) V(\r_i - \r_j) \delta n(\r_j)
	\label{EqHV}
\end{equation}
where the sum $i,j$ runs over all pairs of atomic positions $\r_i$, $\r_j$. We follow estimates for single-layer graphene and interpolate between an onsite repulsion of $V(0)=9.3$ eV and an unscreened $\frac{e^2}{|\r|}$ at large distances. This leads to the following effective form of the screened Coulomb interaction in the tight-binding model,\cite{Wehling:2011cf}
\begin{equation}
	V(\r_i-\r_j) = \frac{1.438}{ 0.116 + |\r_i - \r_j|} \; \mathrm{eV}
	\label{Coulomb}
\end{equation}
where the distance between two carbon atoms $|\r_i - \r_j|$ should be measured in nm.\footnote{Eqn.~(\ref{Coulomb}) is an interpolation between the short-range coupling in Table I of Ref.~[\onlinecite{Wehling:2011cf}], and the long-range result $\epsilon \rightarrow 1$ for $r \rightarrow \infty$ given in the same article.} 

In the Hartree approximation, the product of two density operators in Eqn.~(\ref{EqHV}) is replaced by the product of an operator and its {\em expectation value}. The interaction is now replaced by a site-dependent electric potential $\phi_i$, that is to be determined self-consistently,
\begin{eqnarray}
	H_{H} &=& \sum_i \delta n(\r_i) \phi_i, \label{HamilElectric} \\
	\phi_i &=& \sum_j V(\r_i - \r_j)  \langle \delta n(\r_j) \rangle. \label{Fields}
\end{eqnarray}
This provides us with a self-consistent iterative scheme: using the hopping Hamiltonian -- including the Hartree potential of Eqn.~(\ref{HamilElectric}) -- we compute the band structure, from which we deduce the electronic density on each atom, which allows us to compute the electric potentials $\phi_i$ using Eqn.~(\ref{Fields}). We repeat these steps until we reach convergence, and the fields no longer change.

The band-structure and Bloch wavefunctions were computed on a $8 \times 8$ Monkhorst-Pack momentum grid enlarged to reflect the sixfold rotational symmetry\cite{Monkhorst:1976cv}, which means in total we have 192 momentum points. At each momentum point we compute the full band-structure, and then adjust the chemical potential to find the desired filling. We use mixing of the old and new electric potential to speed up convergence. After 20 iterations the electric potential had converged, in all cases, to within an accuracy of $10^{-4}$ eV. Note that we explicitly exclude lattice symmetry breaking in our computations, in contrast to other works focusing on, for example, $C_3$ breaking.\cite{Choi:2019aa}

We introduce the filling parameter $\nu$, which gives the number of electrons {\em per Moir\'{e} unit cell}, relative to the charge neutrality (where the average charge density {\em per atom} is $\bar{n} = 1$). Note that $\nu = \pm 4$ corresponds to either completely empty or completely filled flat bands, since there are $8$ of them, taking into account spin and valley indices. The average density per atom $\bar{n}$ is related to the filling parameter $\nu$ via $\bar{n} = 1 + \nu / N$, where $N$ is the number of atoms in the unit cell. In the remainder, we will focus on the fillings $\nu = 0, -1, -2, -3$ and $-4$.

\begin{figure*}
	\includegraphics[width=\textwidth]{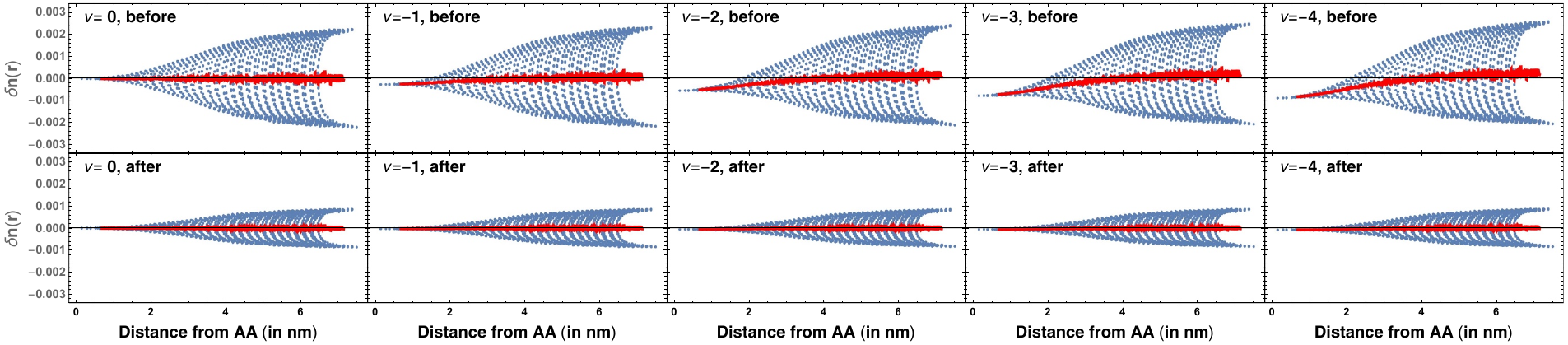}
	\caption{\label{Fig:DensityPlot}
	The relative electron charge density $\delta n (\r)$ as a function of distance from the AA center of the unit cell, for five different fillings ranging from charge neutrality $(\nu=0)$ to the band insulator $(\nu = -4)$.
	The top row shows the charge density without Hartree corrections, the bottom row includes the self-consistent Hartree corrections.
	 The blue dots represent all the atoms in the unit cell. The red line is a moving average, which shows the charge inhomogeneity between the AA center and the AB/BA regions of the unit cell at $\nu<0$. The Hartree correction significantly smoothens the initial charge inhomogeneity.
	}
\end{figure*}

\begin{figure}
	\includegraphics[width=0.49\columnwidth]{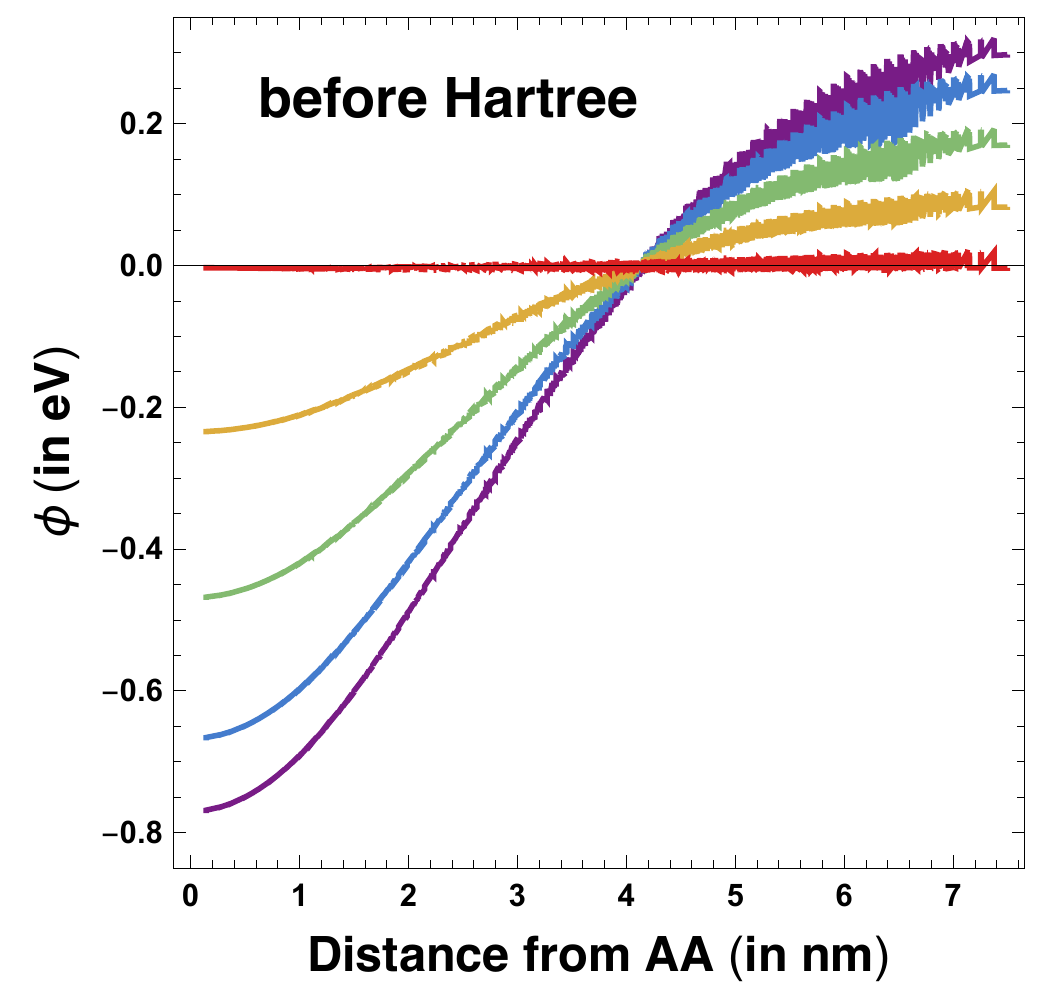}
	\includegraphics[width=0.49\columnwidth]{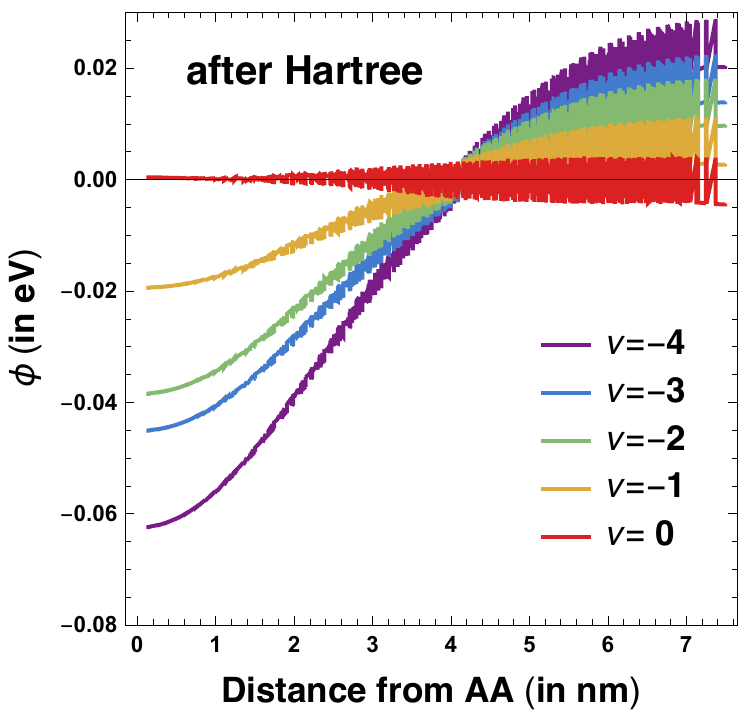}
	\caption{\label{Fig:EPotential}
	The electric potential $\phi_i$ before (left) and after (right) the Hartree corrections as a function of the distance to the AA center, for various fillings. Note the approximate factor 10 reduction in electric potential due to the Hartree corrections.
	}
\end{figure}

\emph{Charge distribution before self-consistency} --
We will first demonstrate the strong inhomogeneity of the charge distribution and electric potential arising in the non-interacting model, without self-consistent Hartree corrections. Interestingly, we find that the inhomogeneity in AB/BA regions is present even at charge neutrality.\footnote{Earlier we assumed that the charge distribution is smooth at charge neutrality,\cite{Rademaker:2018fy} however, this turns out to be wrong}
The distribution of charge at $\nu=0$, illustrated in Fig.~\ref{Fig:ABOscillations}, right, reveals a clear charge deficiency $|\delta n| \lesssim 0.0026$ on the atoms that lie on top of an atom from the other layer. Note that such inhomogeneities exist in AB stacked bilayers only away from charge neutrality.\footnote{A quick calculation of AB stacked bilayers shows that at charge neutrality, the A and B sublattice have the same charge density. Away from charge neutrality, the charge on the sublattices is different.} Furthermore, we find that the AB/BA regions have no net accumulated charge.

Upon hole doping, a significant charge inhomogeneity arises between the AA and the AB/BA regions of the unit cell. This is because of the well-documented fact that the local density of states is highest at the AA region.\cite{TramblydeLaissardiere:2010hla,Luican:2011hw,Wong:2015dg,Guinea:2018kd} The relative charge density is shown in Fig.~\ref{Fig:DensityPlot}, top row. Even though the charge density per atom is $|\delta n| < 0.001$, the combined effect of thousands of atoms leads to a charge accumulation. 
We can express this by a total accumulated charge in the AA region, $\delta n_{\mathrm{AA}}$, obtained by summing the charge over the $N/2=5582$ atoms closest to the AA center. The values for $\delta n_{\mathrm{AA}}$ are shown in Table~\ref{Table:Numbers}.

\begin{table}
	\centering
	\begin{tabular}{||c||c|c||c|c||c||} 
	 \hline
	 $\nu$ & \multicolumn{2}{c}{$\delta n_{\mathrm{AA}}$} 
	 	 &  \multicolumn{2}{c}{$\Delta \phi$ (meV)} & Reduction \\ 
	  & before & after & before & after & factor \\
	 \hline\hline
	 0 & -0.026962 & -0.0049907 & 28.8934 & 8.60985 & 3 - 5\\ 
	 -1& -0.318072& -0.0285706& 339.343 & 30.5505 & 11 \\
	 -2& -0.61295& -0.0515452& 660.412 & 56.6246 & 12 \\
	 -3& -0.87059& -0.0641868& 935.592 & 67.5796 & 14 \\
	 -4& -1.03138 & -0.0850411& 1088.83 & 91.2731 & 12 \\
 	\hline
	 \end{tabular}
	 \caption{\label{Table:Numbers}
	 The total accumulated charge in the AA region, $\delta n_{\mathrm{AA}}$, and the maximum electric potential difference $\Delta \phi$, as a function of the filling $\nu$. We compare the results before, and after the self-consistent Hartree calculation. Both the electric potential and the charge inhomogeneity are severely reduced by the Hartree potential. The reduction factor (before divided by after) is shown in the last column.}
\end{table}

This charge inhomogeneity leads to a very strong electric potential, illustrated in Fig.~\ref{Fig:EPotential}, left. Small potential fluctuations arise because of the charge oscillations in the AB/BA region as shown in Fig.~\ref{Fig:ABOscillations}. At fillings away from charge neutrality, however, a major potential difference arises between the AA center and the AB/BA regions of the unit cell. For filling $\nu=-4$ this difference exceeds 1 eV. A full list of the potential difference $\Delta \phi$ as a function of filling is given in Table~\ref{Table:Numbers}.

\emph{Self-consistent results} --
As shown in Fig.~\ref{Fig:EPotential}, the electric potential is negative around the AA regions. The potential therefore tends to pull more electrons there, thus negating the charge inhomogeneity. Consequently, the Hartree corrections will smoothen the charge distribution within each unit cell. Earlier we qualitatively discussed this phenomenon, which can be viewed as a charge transfer between AA and AB/BA regions.\cite{Rademaker:2018fy} Here we provide a quantitative analysis and discuss its effect on the band-structure.

We find that the Hartree corrections reduce both the charge inhomogeneities and the electric potential by an order of magnitude at fillings other than charge neutrality, as shown in Fig.~\ref{Fig:DensityPlot}, Fig.~\ref{Fig:EPotential} and Table~\ref{Table:Numbers}.  

\begin{figure}
	\includegraphics[width=\columnwidth]{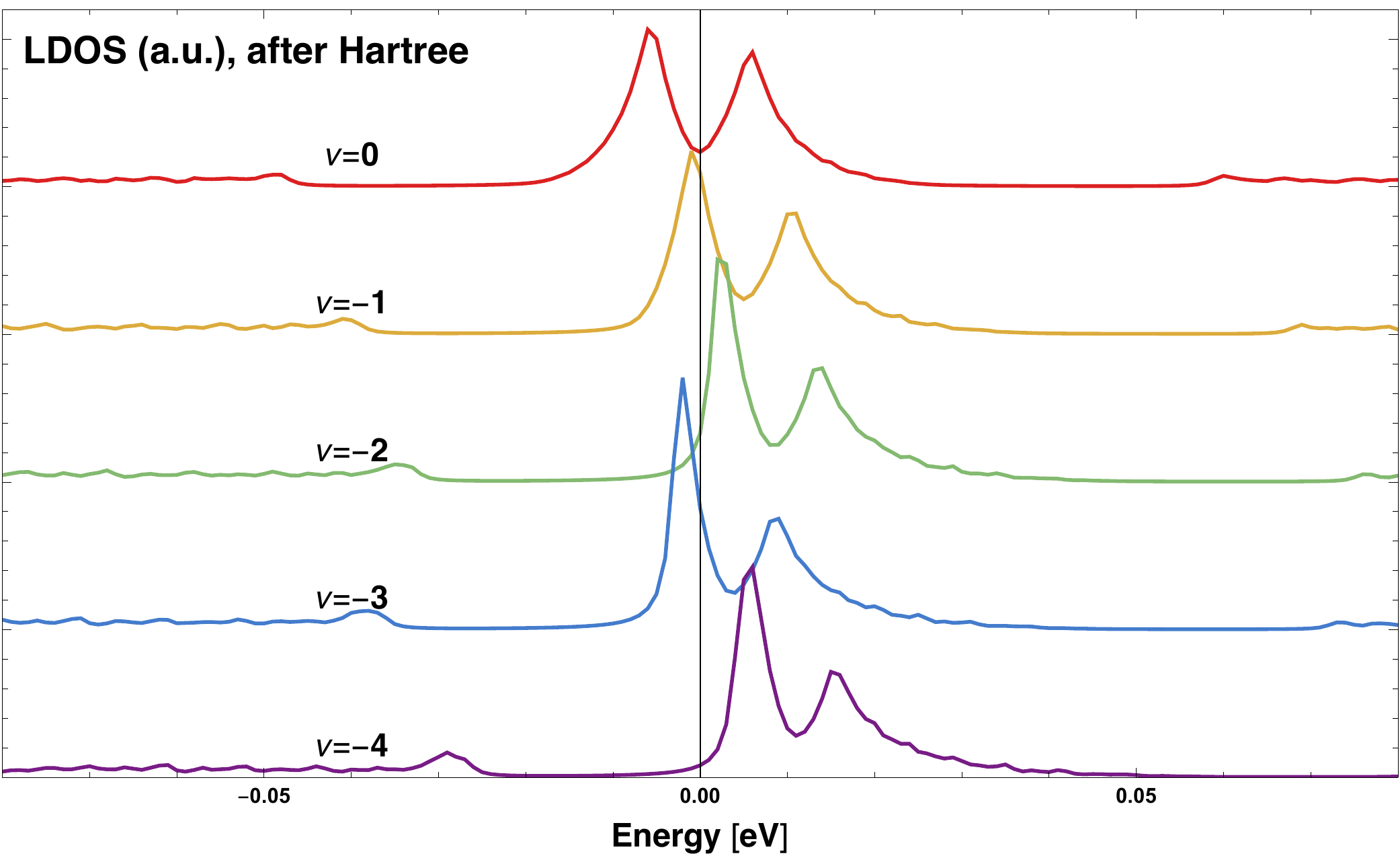}
	\caption{\label{Fig:LDOS}
	The local density of states measured at the AA region of the unit cell, in arbitrary units, for fillings $\nu = 0$ through $\nu = -4$. At charge-neutrality ($\nu = 0$) the two Van Hove singularities are clearly separated. For partial fillings of the flat band, the Van Hove singularity remains close to the Fermi level. A similar effect that has been observed in Ref.~\cite{Cea:2019tb}.
	}
\end{figure}

\begin{figure*}
	\includegraphics[width=\textwidth]{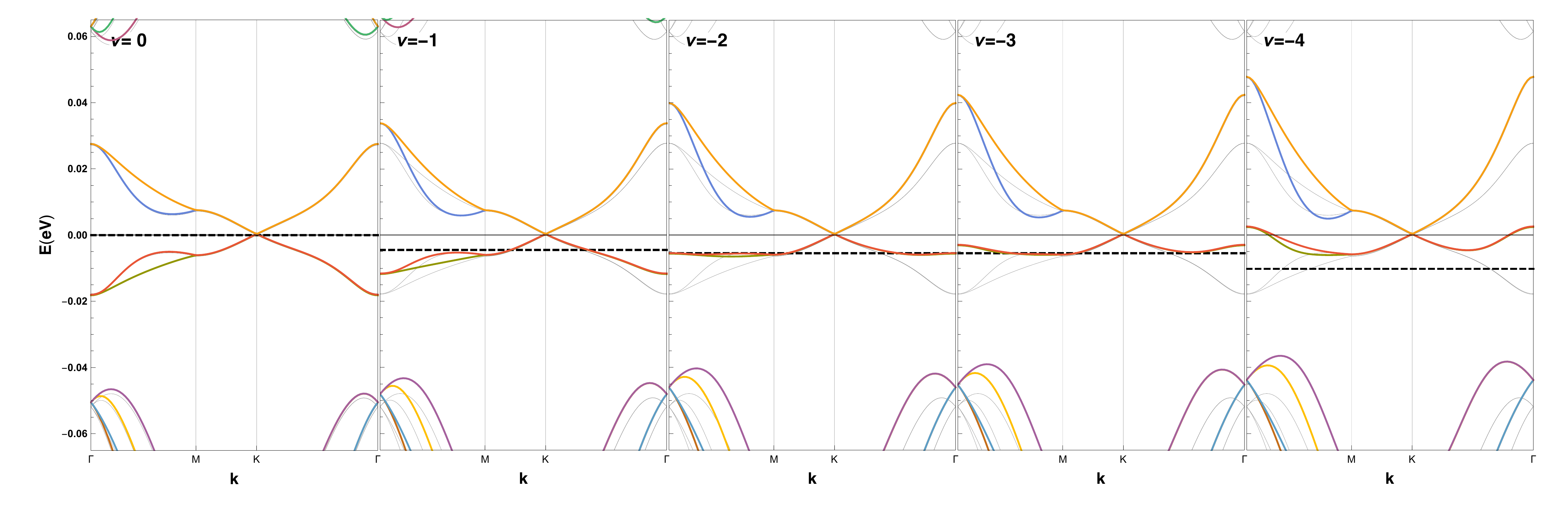}
	\caption{\label{Fig:BandStructure}
	The band structure after Hartree corrections are taken into account, for fillings from $\nu=0$ to $\nu=-4$. The thin grey lines represent the band structure prior to the Hartree corrections. The thick dashed line is the Fermi level at each given filling. Upon doping away from charge neutrality, the states at ${\bf \Gamma}$ are clearly pushed upwards, which is a feature of the charge smoothening. For the flat bands, this implies a reduction of the bandwidth.
	}
\end{figure*}


The charge redistribution affects dispersion of the bands, as well as the local density of states (LDOS).
In Fig.~\ref{Fig:LDOS} we show the LDOS, measured at the AA region of the unit cell, for various fillings. At charge neutrality there are symmetric peaks representing the Van Hove singularities (VHS). When partially filling the flat bands (here the fillings $\nu=-1, -2$ and $\nu=-3$) the VHS remains close to the Fermi level, consistent with recent findings a continuum model.\cite{Cea:2019tb}

The change of position of the VHS is a direct consequence of band structure changes. Recall that in the non-interacting picture the Bloch wavefunctions at the ${\bf K}$ point are highly centered at the AA regions of the unit cell ({\em center} orbitals),\cite{Guinea:2018kd,Rademaker:2018fy}. On the other hand, the real-space structure of the flat bands at the ${\bf \Gamma}$ point has a vanishing weight at the AA centers and instead resembles a {\em ring} occupying the AB/BA regions. Charge smoothening can be achieved by transferring charge from the center to the ring orbitals. 
Consequently, for hole doping away from charge neutrality ($\nu < 0$), the charge smoothening will push up the energy of the band at the ${\bf \Gamma}$ point. This can be clearly seen in Fig.~\ref{Fig:BandStructure}, where we show the band structure after including the Hartree corrections. 

The states at the ${\bf \Gamma}$ point are clearly pushed upwards, to the point that at $\nu=-4$ they are higher in energy than the states at the ${\bf K}$ point. This is true both for the flat bands as well as the bands lower in energy. However, for the flat bands this upwards push directly causes an even further flattening, in the form of a reduction of the bandwidth from $18\,{\rm meV}$ to roughly $6\,{\rm meV}$ at its extreme. Especially noteworthy is the filling $\nu=-2$, where the charge smoothening leads to the appearance of an almost completely flat band segment between the ${\bf \Gamma}$ and ${\bf M}$ points. 
It will be interesting to investigate the effect of this significant flattening on the correlated insulator state observed at this filling.\cite{Cao:2019ui,Cao:2018kn,Yankowitz:2018tx}


\emph{Discussion} --
We have shown that Coulomb interactions, as captured by a fully self-consistent Hartree calculation, smoothen the charge inhomogeneities in the unit cell. This charge smoothening is achieved by pushing up the energy of the states at the ${\bf \Gamma}$ point, which in turn forces the VHS to remain close the Fermi level.

Note that in this work we assumed an interaction strength similar to those expected for single-layer graphene. However, both screening within the bilayer as well as the dielectric environment\cite{Pizarro:2019wf} can alter the precise electron-electron interactions. Nevertheless, a different shape of the Coulomb interaction will only quantitatively change our results as its tendency to smooth out charge inhomogeneities remains. Similar quantitative changes are expected when one includes the lattice relaxation response to the Hartree potential, which is not included in this work.

Both the redistribution of charge and the shift in VHS position upon doping can, in principle, be observed using scanning tunnelling microscopy (STM). Indeed, a sharpening of the VHS closest to the Fermi level has been observed,\cite{Choi:2019aa,Xie:2019vz,Jiang:2019to} but further many-body effects, not considered here, make a quantitative comparison between our results and the observed STM spectra difficult.

Indeed, to quantitatively understand the plethora of interesting observed phases, ranging from superconductivity, correlated insulator states, quantum anomalous Hall effect, ferromagnetism, requires the inclusion of many-body interactions beyond Hartree corrections.\cite{Cao:2018kn,Cao:2018ff,Cao:2019ui,Yankowitz:2018tx,Choi:2019aa,Xie:2019vz,Jiang:2019to,Lu:2019ww,Sharpe:vc,Ochi:2018ug,Kang:2018tt} 
This, in turn, demands the computation of localized Wannier orbitals and their Hubbard, exchange and Hund couplings. Given the strong band flattening and charge smoothening the Hartree correction gives, we argue that any effective low-energy model should start from a Hartree-renormalized band structure such as the one shown in Fig.~\ref{Fig:BandStructure}.

{\it Note added.} Upon finalization of this manuscript, a related paper appeared.\cite{Cea:2019tb} This work starts from a continuum model without lattice relaxation, while we start from a tight-binding model with lattice relaxation. Otherwise, the resulting band-structure and LDOS spectra are qualitatively similar. A major difference with Ref.~[\onlinecite{Cea:2019tb}] is that we discuss the renormalization of the band structure in terms of the charge inhomogeneities and the resulting electric fields, while authors of Ref.~[\onlinecite{Cea:2019tb}] interpret their results in terms of VHS pinning to the Fermi level.

\acknowledgments \emph{Acknowledgments} -- We are thankful to Oleg Yazyev for sharing his relaxed lattice results.
This work is supported by the Swiss National Science Foundation via an Ambizione grant (L.~R.), and via a regular project (D.~A.), and by the Fondecyt Grant No. 1160239 (P.~M.).


\end{document}